\documentclass[11pt]{article}       

\usepackage{graphicx}
\usepackage{amsmath}
\usepackage{amssymb}

\usepackage[letterpaper,margin=4cm]{geometry}

\begin{document}
\newcommand{\ket}[1]{\left|{#1}\right>}
\newcommand{\bra}[1]{\left<{#1}\right|}
\newcommand{\ketbra}[2]{\left|{#1}\right>\left<{#2}\right|}
\newcommand{\braket}[2]{\left<{#1}\right|\left.{#2}\right>}
\newcommand{\cL}{{\mathcal L}}
\newcommand{\cK}{{\mathcal K}}
\newcommand{\cP}{{\mathcal P}}
\newcommand{\cQ}{{\mathcal Q}}
\newcommand{\Tr}{{\rm Tr}}
\renewcommand{\arraystretch}{1.5}

\title{\bf{
The Dirac-Moshinsky 
oscillator coupled to an external field 
and its connection to quantum optics}
}

\author{ 
Juan Mauricio Torres$^*$,  
Emerson Sadurn\'i$^\dagger$ 
and 
Thomas H. Seligman$^{*,**}$
\\ 
\vspace{-.3cm}
\small{\it{ $^*$Instituto de Ciencias F\'isicas, Universidad Nacional 
Aut\'onoma de M\'exico,} }
\\
\vspace{-.3cm}
\small{\it{  Cuernavaca, Morelos, 62210, M\'exico}},
\\
\vspace{-.3cm}
\small{\it{ $^\dagger$Universit\"at Ulm, Institut f\"ur Quantenphysik, D-89069 Ulm, Germany}}
\\
\vspace{-.3cm}
\small{\it{ $^{**}$Centro Internacional de Ciencias, Cuernavaca, Morelos, M\'exico}}
 }
 \date{}

\maketitle

\begin{abstract}
  The Dirac-Moshinsky oscillator is an elegant example of an 
exactly solvable quantum relativistic model that
  under certain circumstances can be mapped onto the Jaynes-Cummings model in quantum optics. 
  In this work we show, how to do this in detail. Then we extend it by considering its 
  coupling with an external (isospin) field and find the 
conditions that maintain solvability. 
  We use this extended system to explore entanglement in relativistic systems and then
  identify its quantum optical analog: two different atoms interacting with an electromagnetic mode.
  We show different aspects of entanglement which gain relevance in this last system,
which can be used to emulate the former.
\end{abstract}

\section{Introduction}
The fact that the Dirac equation is analytically solvable both for the free particle 
and the Coulomb problem made it plausible to look for other solvable Dirac 
systems whose non-relativistic counterparts, in analogy to the above, 
are superintegrable and/or algebraically solvable.
Indeed additional problems are solvable in the relativistic 
setting; in particular, a generalization of the harmonic oscillator, 
with its large dynamical symmetry
\cite{marcosbook}. Moshinsky and Szczepaniak \cite{mosh} did not 
only realize this, but constructed explicit solutions. This line was extensively followed
by Moshinsky and his collaborators \cite{mosh,lista}.

We shall show that this solvability results from an additional conserved quantum number.
This conservation indeed allows to preserve solvability in an algebraic sense even as we 
couple an additional isospin field into the system. Appropriate coupling conserves the 
quantum number and thus the symmetry of the extended problem.
This solution was clarified by a mapping onto a more general 
problem in quantum optics, which we solved recently \cite{torres}.

The connection to the quantum optics problem not only makes the solvability more 
understandable in a wider setting but also allows to conceive quantum optics experiments 
that emulate this system. Nowadays these experiments are feasible with
trapped ions. The free particle
Dirac equation has already been emulated 
\cite{lamata,blatt} and the mapping of the 
standard Dirac-Moshinsky oscillator (DMO) in $2+1$ dimensions on the 
Jaynes-Cummings model (JCM) has been proposed \cite{delgado}.
Here  we shall review these methods and also show some cases in which 
the $3+1$ dimensional example can be mapped.
We then propose a system of two atoms interacting with a cavity with 
different couplings on which the DMO coupled to an isospin field
 can in principle be realized.

The paper is organized as follows. In the section \ref{sec:DMO} we 
review and describe the Dirac-Moshinsky oscillator, in section \ref{sec:JC} we
present the Jaynes-Cummings model which is of great importance in quantum optics
and cavity QED. In section \ref{sec:DMOmap} we describe how to map
the relativistic model of the DMO onto the JCM for the three dimensional cases,
this will allow us to formulate a straightforward method to extend
the model as presented in section \ref{sec:DMOcoup}. In section
\ref{sec:cav2at} the quantum optical model related to the extended model
will be presented. 

\section{The Dirac-Moshinsky oscillator}\label{sec:DMO}
The DMO, introduced in 1989 by Marcos Moshinsky and A.
Szczepaniak is a solvable quantum relativistic model which in the non-relativistic
limit  corresponds to the harmonic oscillator plus a  spin-orbit coupling term. 
Noting that the momentum operator is linear in the Dirac
equation, they proposed a linear term in position as well.  We chose to write it as
follows:
\begin{equation}\label{eq:diraceq}
  i\hbar \frac{\partial \ket{\Psi}}{\partial t}=
  \left(
  c\boldsymbol{\alpha}\cdot\left(\boldsymbol{p}+im\omega\beta\boldsymbol{r}\right)
  +mc^2\beta\right)\ket{\Psi}=H\ket{\Psi}.
\end{equation}
Here $c$ denotes  the speed of light, $m$ and $\omega$
are the mass and frequency of the oscillator, 
and we make use of the following Dirac matrices: 
\begin{equation}\label{eq:diracmat}
  \boldsymbol{\alpha}=-\sigma_y\otimes\boldsymbol{s},
  \qquad
  \beta=\sigma_z\otimes\mathbb{I}_2.
\end{equation}
We use two sets of Pauli matrices, namely $\boldsymbol{s}=(s_x,s_y,s_z)$
for the spin of the particle and $\boldsymbol{\sigma}=(\sigma_x,\sigma_y,\sigma_z)$
for the isospin.
The representation of both $\boldsymbol{\sigma}$ and $\boldsymbol{s}$
is the same, but we use different symbols to avoid confusion. 
We also use the convention of writing the isospin terms always at the left and
in the following we shall omit to write the outer product $\otimes$ explicitly 
to simplify notation. 

Using the raising and  lowering operators 
$\sigma_\pm=\tfrac{1}{2}(\sigma_x\pm i\sigma_y)$
and the isospin states $\ket{\pm}=\sigma_{\pm}\ket{\mp}$ 
the state vector can be expressed as
\begin{equation}\label{eq:satevec}
  \ket{\Psi}
  =
  \ket{-}\ket{\psi_1}+
  \ket{+}\ket{\psi_2}
\end{equation}
and the Hamiltonian can be rewritten as  
\begin{equation}\label{eq:dhamalgrp}
  H=
  mc^2\sigma_z
  +c\, 
  \sigma_-\boldsymbol{s}\cdot\left(m\omega\boldsymbol{r}+i\boldsymbol{p}\right)
  +c\, 
  \sigma_+\boldsymbol{s}\cdot\left(m\omega\boldsymbol{r}-i\boldsymbol{p}\right).
\end{equation}
This form shows clearly  the coupling between
$\ket{\psi_1}$ and 
$\ket{\psi_2}$, the big and small component of the state vector respectively.  Squaring the Hamiltonian
results in
\begin{equation}\label{eq:squareham}
\frac{E^2-m^2c^4}{c^2}\ket{\psi_1}= 
\left(
p^2+m^2\omega^2r^2-3\hbar\omega m c^2-2m c^2\omega\boldsymbol{s}\cdot\boldsymbol{L}
\right)\ket{\psi_1},
\end{equation}  
where we have introduced the angular momentum operator   
$\boldsymbol{L}=\boldsymbol{r}\times\boldsymbol{p}$.
An analogous expression can be obtained for $\ket{\psi_2}$.
At this point one can take the non relativistic limit 
noting that $E=mc^2+\varepsilon$, with $\varepsilon\ll mc^2$ being
the non-relativistic energy.  The term at the left part 
becomes approximately $2mc^2\varepsilon$, 
which means that
$\varepsilon$ is an eigenvalue of the operator at the right hand side
which we recognize easily as the Hamiltonian of an isotropic harmonic
oscillator plus a constant and a spin-orbit coupling term.

From equation \eqref{eq:squareham} it is evident that $H^2$ 
commutes with
$\boldsymbol{J}^2$ ($\boldsymbol{J}=\boldsymbol{L}+\boldsymbol{S}$), the total
angular momentum. 
For $H$ one has an additional constant of motion, 
$\boldsymbol{a^\dagger}\cdot\boldsymbol{a}+\frac{1}{2}\sigma_z$
that becomes clearer if one express the Hamiltonian in terms
of the oscillator  creation an annihilation operators 
$a=\sqrt{\tfrac{m\omega}{2\hbar}}(\boldsymbol{r}+i\tfrac{\boldsymbol{p}}{m\omega})$ and
$a^\dagger=\sqrt{\tfrac{m\omega}{2\hbar}}(\boldsymbol{r}-i\tfrac{\boldsymbol{p}}{m\omega})$
\begin{equation}\label{eq:diracohamalg}
  H=
  mc^2\sigma_z
  +\eta
  \left(
  \sigma_-\boldsymbol{s}\cdot\boldsymbol{a^\dagger}+
  \sigma_+\boldsymbol{s}\cdot\boldsymbol{a}
  \right),
\end{equation}
where we have defined $\eta=\sqrt{2mc^2\hbar\omega}$. 
This means that the eigenfunctions of $H$  can
be expressed as expressed as a combination
of $\ket{\pm}$, the isospin components,  
and $\ket{n\left(j\pm\frac{1}{2},\frac{1}{2}\right)jm_j}$, the
eigenstates of $3{\rm D}$ harmonic oscillator coupled to spin $\tfrac{1}{2}$.
$n$ is the radial quantum number, $j$ is the angular momentum quantum number 
and $m_j$ its projection.

Before going into any further details
of the DMO eigensystem, we take a short detour to introduce
the quantum optical model related to our construction. 
\section{The Jaynes-Cummings model}\label{sec:JC}
The JCM \cite{bunch} is a paradigm in quantum optics which
can be thought of as a model that 
describes the interaction between a two level atom
and one mode of the electromagnetic field, {\it i. e.} a cavity mode.
The Hamiltonian describing such system can be written as
\begin{equation}\label{eq:jcham}
  H_{{\rm JC}}=\Omega(\sigma_+a+\sigma_+a^\dagger)+\delta\sigma_z,
\end{equation}
where $a^\dagger$ and $a$ represent the creation and annihilation
operator of photons in the cavity and $\sigma_{\pm}$ are the raising
and lowering operators of the atomic states. 
$\Omega$
represents the coupling strength between
the cavity and the atom and $\delta$ stands for the detuning of the atomic transition
frequency from the cavity mode.
The energy term of the mode is absent from \eqref{eq:jcham},
because it is written in the interaction picture.

To diagonalize the Hamiltonian \eqref{eq:jcham} one notes
that there is a conserved quantity
$I=a^\dagger a+\tfrac{1}{2}\sigma_z$ which can be
identified as the number of excitations in the system as
there is only excitation exchange in the system.
Using the basis
$\ket{-}\ket{\mathfrak{n}}$, $\ket{+}\ket{\mathfrak{n}-1}$,
where $\ket{-}$, $\ket{+}$ are the ground and excited state of 
the atom and $\ket{\mathfrak{n}}$ is a number state in the cavity,
the Hamiltonian is 
block-diagonal in terms of $2\times 2$ matrices:
\begin{equation}\label{eq:matjc}
  H_{{\rm JC}}(\mathfrak{n})=\left(
  \begin{array}{cc}
    \delta&\Omega\sqrt{\mathfrak{n}}\\
    \Omega\sqrt{\mathfrak{n}}&-\delta
  \end{array}
  \right).
\end{equation}
The eigenenergies can be easily obtained:
\begin{equation}
  \mathcal{E}_{\pm}(\mathfrak{n})=\pm\sqrt{\delta^2+\Omega^2\mathfrak{n}}=\pm\mathcal{E}(\mathfrak{n})
\end{equation}
and one can find the corresponding eigenstates, which are known as
dressed states in the literature:
\begin{align}\label{eq:dress}
  \ket{\varphi_{+}(\mathfrak{n})}&
  =\sin{(\theta_\mathfrak{n})}\ket{-}\ket{\mathfrak{n}}
  +\cos{(\theta_n)}\ket{+}\ket{\mathfrak{n}-1}\nonumber\\
  \ket{\varphi_{-}(\mathfrak{n})}&
  =\cos{(\theta_\mathfrak{n})}\ket{-}\ket{\mathfrak{n}}
  -\sin{(\theta_n)}\ket{+}\ket{\mathfrak{n}-1}
\end{align}
with 
\begin{equation}\label{eq:thetajc}
\theta_\mathfrak{n}=\arctan{\left(
\sqrt{\frac{\mathcal{E}(\mathfrak{n})-\delta}{\mathcal{E}(\mathfrak{n})+\delta}}\right)}
\end{equation}
Now that we have fixed the notation for the JCM 
and shown the simplicity of its solutions, we proceed to relate it with the DMO 
in the next section.

\section{Mapping the DMO onto the JCM}\label{sec:DMOmap}
In this section we describe the connection between the DMO 
and the JCM and show under which circumstances
the DMO can be mapped onto the JCM. 
\subsection{$1+1$ DMO}
Let us consider only one spatial dimension, namely the $1+1$ DMO.
For this case one needs only two anticommuting Dirac matrices and we choose to write
\begin{equation}
  H^{(1)}=-c\sigma_y(p+im\omega\sigma_z x)+mc^2\sigma_z.
\end{equation}
where the superscript in the Hamiltonian indicates that we are working
in the one-dimensional case.
Using the creation and annihilation operators 
$a_x^\dagger=\sqrt{\frac{m\omega}{2\hbar}}x-i\frac{p}{m\omega}$,
$a_x=\sqrt{\frac{m\omega}{2\hbar}}x+i\frac{p}{m\omega}$
and the raising and lowering operators
$\sigma_\pm=(\sigma_x\pm\sigma_y)/2$ of the Dirac Spinor
one can rewrite the previous equation as
\begin{equation}
  H^{(1)}=\sqrt{2mc^2\hbar\omega}
  \left(\sigma_+a_x+\sigma_-a_x^\dagger\right)
  +mc^2\sigma_z.
\end{equation}
If one takes a look at equation \eqref{eq:jcham}, the connection in this case is obvious.
This Hamiltonian is exactly the JCM Hamiltonian in quantum optics.
Thus the $1+1$ DMO maps exactly onto
the JCM, provided  that one identifies 
$\sqrt{2mc^2\hbar\omega}\to\Omega$,
 $mc^2\to\delta$,
the isospin with the atomic system
and the spatial degrees of freedom with the cavity mode.

\subsection{$2+1$ DMO}

Now let us consider the case in two spatial dimensions, the $2+1$ DMO. For a full
description of this case see \cite{delgado,delgadochiral}.
Here we need three anticommuting Dirac matrices and we choose
\begin{equation}
  H^{(2)}=-c\sigma_x(p_y+imc^2\sigma_z y)-c\sigma_y(p_x+imc^2\sigma_z x)+mc^2\sigma_z
\end{equation}
The ladder operators for each spatial dimension $x$ and $y$ can be used to   
construct a chiral representation in the form
\begin{align}
a_l=(a_x-ia_y)/\sqrt{2}\qquad a_l^\dagger=(a_x^\dagger+ia_y^\dagger)/\sqrt{2}
\nonumber\\
a_r=(a_x+ia_y)/\sqrt{2}\qquad a_r^\dagger=(a_x^\dagger-ia_y^\dagger)/\sqrt{2}
\end{align}
which are also creation and annihilation operators with 
the canonical commutation rule $[a_r,a_r^\dagger]=1$ and $[a_l,a_l^\dagger]=1$.
Using the chiral ladder operators, together with the previously 
defined $\sigma_{\pm}$, one finds
\begin{equation}\label{eq:DMOh2}
  H^{(2)}=2\sqrt{mc^2\hbar\omega} 
  \left(\sigma_+a_r+\sigma_+a_r^\dagger\right)+mc^2\sigma_z.
\end{equation}
The set of operators $a_l$, $a_l^\dagger$ is absent from equation \eqref{eq:DMOh2},
which means that the eigenstates of $H^{(2)}$ depend only on
the number states $\ket{n_r}$, of the number operator $a_r^\dagger a_r$, and are infinitely degenerate
in the subspace spanned by $\ket{n_l}$.
The connection to the JCM is also obvious. 
One has to identify the subspace
of $a_r$ with the cavity mode, the isospin with the atomic system
and in this case $mc^2\to\delta$ and $2\sqrt{mc^2\hbar\omega}\to\Omega$.

\subsection{3+1 DMO}
Returning to the $3+1$ case described by  the Hamiltonian 
in equation \eqref{eq:diracohamalg},
one recognizes that 
$I^{(3)}=\boldsymbol{a}^\dagger\cdot\boldsymbol{a}+\tfrac{1}{2}\sigma_z$
is a conserved quantity. This tells us that a natural
way of labeling the 
eigenstates of the total angular momentum 
and the 3D harmonic oscillator,
is in terms of the oscillator quantum number $N=2n+j\pm\tfrac{1}{2}$.
We define
\begin{equation}\label{eq:ketN}
  \ket{N}=\ket{n\left(j\pm\tfrac{1}{2},\tfrac{1}{2}\right)jm_j}.
\end{equation}
There is no ambiguity in the previous definition if one remembers the dependence
of $N$ on both $j$ and $n$, the total angular momentum and radial oscillator
quantum numbers. The key here is the parity of $N$ as there
are two orthogonal states with the same value of $j$ and $n$ ,
in eq. \eqref{eq:ketN}, each one
of these will be labeled by an 
$N$ with the same parity as $j\pm\tfrac{1}{2}$. The lowest
possible value will be $N_{{\rm min}}=j-\tfrac{1}{2}$.

In order to find the eigenstates of $H$
one has to know how the ladder operator $\boldsymbol{s}\cdot\boldsymbol{a}$, and its
hermitian conjugates, act on the number states we just defined.
We take the result from  reference \cite{marcosbook} and write it as
\begin{align}\label{eq:ladders3simpl}
  \boldsymbol{s}\cdot\boldsymbol{a}\ket{N}&=\sqrt{\mu(N)}\ket{N-1}\nonumber\\
  \boldsymbol{s}\cdot\boldsymbol{a}^\dagger\ket{N-1}&=\sqrt{\mu(N)}\ket{N},
\end{align}
with
\begin{equation}
  \mu(N)=\left\{
  \begin{array}{c}
    \sqrt{2n+2j+2}\qquad N=2n+j+\tfrac{1}{2}\\
    \sqrt{2n}\qquad N=2n+j-\tfrac{1}{2}.
\end{array}\right.
\end{equation}
which takes into account the two separate cases, when $N$ has the parity of
$j+\tfrac{1}{2}$ and $j-\tfrac{1}{2}$.
Using the   
basis where $I^{(3)}$
is diagonal, namely
$\ket{-}\ket{N}$ and $\ket{+}\ket{N-1}$, one can 
diagonalize the Hamiltonian in equation 
\eqref{eq:diracohamalg}
in terms of $2\times2$ matrices
\begin{equation}\label{eq:HN}
  H(N)=\left(
  \begin{array}{cc}
    mc^2&\eta\sqrt{\mu(N)}\\
    \eta\sqrt{\mu(N)}&-mc^2
  \end{array}
  \right).
\end{equation}
In complete analogy with the
JCM (see equation \eqref{eq:matjc}) one can express the 
corresponding eigenergies in the following form:
\begin{equation}\label{eq:EN}
  E_{\pm}(N)=\pm\sqrt{m^2c^4+\eta^2\mu(N)}=\pm E(N).
\end{equation}
The eigenstates will have the same functional form as the dressed states
of the JCM in equation 
\eqref{eq:dress}, but with the number states $\ket{N}$ and energies $E(N)$.

The eigenstates with $N=2n-j+\tfrac{1}{2}$ are infinitely degenerate as 
they have energies that do not depend on $j$ as can be seen from \eqref{eq:EN}.
In addition, 
the blocks in equation \eqref{eq:HN} have the same form to those
of the JCM in equation \eqref{eq:matjc}. 
Therefore we can state that the infinitely degenerate part of the  $3+1$ DMO
can be mapped to the JCM,
if one identifies $\sqrt{2}\eta\to\Omega$ and $mc^2\to\delta$.

If one takes $N=2n+j+\tfrac{1}{2}$ the degeneracy is finite. In this case
the functional dependence of the blocks in equation \eqref{eq:HN} with $N$ is 
different to the JCM, so it can not be fully mapped. Even so, one could
still emulate the finite degenerate part of the DMO in a JCM
if one restricts only to one $2\times 2$ block of $H^{(N)}$, here one should 
identify $\eta\sqrt{2n+2j+2}$ with $\Omega\sqrt\mathfrak{n}$.


\section{A DMO coupled to an external field}\label{sec:DMOcoup}
Now we present the extension to the DMO 
interacting with an isospin field  modeled as a potential that is summed to
the total Hamiltonian
\begin{equation}
  \tilde H= H+\Phi.
\end{equation}
Among the many choices which preserve the integrability of the system, 
we use the simplest (i.e. linear) one by way of example, namely
\begin{equation}
  \Phi=\chi(\sigma'_-\mathcal{A}^\dagger+\sigma'_+\mathcal{A})+\gamma \sigma'_z,
\end{equation}
Here $\mathcal{A}$ represents the ladder operator
for each dimensionality we considered in the previous section, and we have denoted 
with primes the isospin operator that
acts on the field degrees of freedom.
The full Hamiltonian is given by
\begin{equation}\label{eq:hamDOMc}
\tilde H=
\eta(\sigma_-\mathcal{A}^\dagger+\sigma_+\mathcal{A})+
\chi(\sigma'_-\mathcal{A}^\dagger+\sigma'_+\mathcal{A})
+mc^2\sigma_z+\gamma \sigma'_z.
\end{equation}
For a physical discussion on $\Phi$ and the covariant form of 
this system see \cite{emerson}.

We shall proceed from here taking $\mathcal{A}$ as any ladder operator,
that satisfies
\begin{equation}
 \mathcal{A}\ket{\mathcal{N}}=f(\mathcal{N})\ket{\mathcal{N}-1}.
\end{equation}
Next we note that, due to the additional isospin,
one has the integral of motion\footnote{It will be further defined
for each dimensionality considered}
\begin{equation}
  I=\mathcal{A}^\dagger\mathcal{A}+\tfrac{1}{2}\left(\sigma_z+\sigma_z'\right).
\end{equation}
Using the basis where $I$ is diagonal, namely
\begin{equation}\label{eq:extbasis}
  \ket{-}\ket{-'}\ket{\mathcal{N}+1}\quad 
  \ket{+}\ket{-'}\ket{\mathcal{N}}\quad 
  \ket{-}\ket{+'}\ket{\mathcal{N}}\quad 
  \ket{+}\ket{+'}\ket{\mathcal{N}-1},
\end{equation}
the Hamiltonian is now block diagonal with its blocks given by the $4\times4$ matrices
\begin{equation}\label{eq:matDMOc}
\tilde H(\mathcal{N})=
\left(
\begin{array}{cccc}
-mc^2-\gamma& \chi f(\mathcal{N}+1)& \eta f(\mathcal{N}+1)&0\\
\chi f(\mathcal{N}+1)&\gamma-mc^2&0& \eta f(\mathcal{N})\\
\eta f(\mathcal{N}+1)&0&mc^2-\gamma& \chi f(\mathcal{N})\\
0& \eta f(\mathcal{N})& \chi f(\mathcal{N})&mc^2+\gamma
\end{array}
\right),
\end{equation}
where $f(\mathcal{N})$ depends on the dimensionality one choses.

The analysis in the previous section has allowed us to construct 
the generalization of DMO oscillator, in any of the three 
dimensionalities considered, coupled to
an external isospin field.  Table \ref{tab:coup} shows for each dimensionality 
the correspondence of the ladder  operator $\mathcal{A}$, 
the integral of motion $I$,  the quantum number $\mathcal{N}$ and the function $f$ introduced
in equation \eqref{eq:matDMOc}.

\begin{table}[h!]
\begin{center}
\label{tab:coup}
\caption{Correspondence for each dimensionality.}
\begin{tabular}{|l|l|l|l|l|}
\cline{1-5}
DMO&$\mathcal{A}$&Conserved quantum number $I$&$f(\mathcal{N})$&$\ket{\mathcal{N}}$\\
\cline{1-5}
$1+1$&$a_x$&$a_x^\dagger a_x +\tfrac{1}{2}(\sigma_z+\sigma'_z)$&
$\sqrt{n}$&$\ket{n}$\\
\cline{1-5}
$2+1$&$\sqrt{2}a_r$&$a_r^\dagger a_r +\tfrac{1}{2}(\sigma_z+\sigma'_z)$&
$\sqrt{2}\sqrt{n_r}$&$\ket{n_r}$\\
\cline{1-5}
$3+1$&$\boldsymbol{s}\cdot\boldsymbol{a}$&
$\boldsymbol{a^\dagger}\cdot\boldsymbol{a} +\tfrac{1}{2}(\sigma_z+\sigma'_z)$&
$\sqrt{\mu(N)}$&$\ket{N}$
\\
\cline{1-5}
\end{tabular}
\end{center}
\end{table}
The systems are again integrable and one can find the eigenenergies by 
diagonalizing each block $\tilde H(\mathcal{N})$.
We shall not write the general solutions here as they can be found in \cite{emerson}.
Instead consider
the evolution of a simple initial state and evaluate the entanglement of the DMO with the field.

\subsection{Entanglement with the field}\label{subsec:ef}
In this section we analyze the dynamical features of a 
Dirac particle under the influence of the external field. To this end and for 
simplicity, we
use a product initial state formed by the lowest eigenstate of the DMO times 
the upper state of the field, namely
\begin{equation}\label{eq:DMOcist}
  \ket{\Psi_0}=\ket{-}\ket{+'}\ket{0}
\end{equation}
With our choice $\mathcal{N}=0$ it follows from equation \eqref{eq:extbasis}
that the basis is reduced to three states, because $\mathcal{N}$ has to be non-negative.
Thus the evolution will stay confined in a $3$ dimensional 
subspace\footnote{For $\mathcal{N}>0$ one would have $4$ dimensional subspaces.}
of the entire Hilbert space and the state vector at any time
can be written as
\begin{align}\label{eq:DMOctst}
  \ket{\Psi(t)}=
  B_1(t)\ket{-}\ket{-'}\ket{1} +
  B_2(t)\ket{+}\ket{-'}\ket{0} 
  +B_3(t)\ket{-}\ket{+'}\ket{0}.
\end{align} 
If one simplifies things even more, by setting $\eta=\chi$ and $mc^2=\gamma$
one can find simple explicit solutions for these coefficients, namely
\begin{align}
B_1(t)=&f_0(t)\nonumber\\
B_l(t)=&\tfrac{1}{2}\left(1-f_0(t)+(-1)^l g(t)\right)
\quad l=2,3
\end{align}
with the definitions
\begin{align}\label{eq:fg}
  \tilde\gamma=&\sqrt{\gamma^2+2}\nonumber\\
  f_0(t)=&\frac{1}{\tilde\gamma^2}\sin{\left(t\tilde\gamma\right)}
  \nonumber\\
  g(t)=&
  \frac{\gamma+\tilde\gamma}{2 \tilde\gamma}
  \cos{\left(t\left(\tilde\gamma-\gamma\right)\right)}+
  \frac{1}{\tilde\gamma(\gamma+ \tilde\gamma)}\cos{\left(t\left(\tilde\gamma+\gamma\right)\right)}.
\end{align}
\begin{figure}[t!]
  \includegraphics[width=\textwidth]{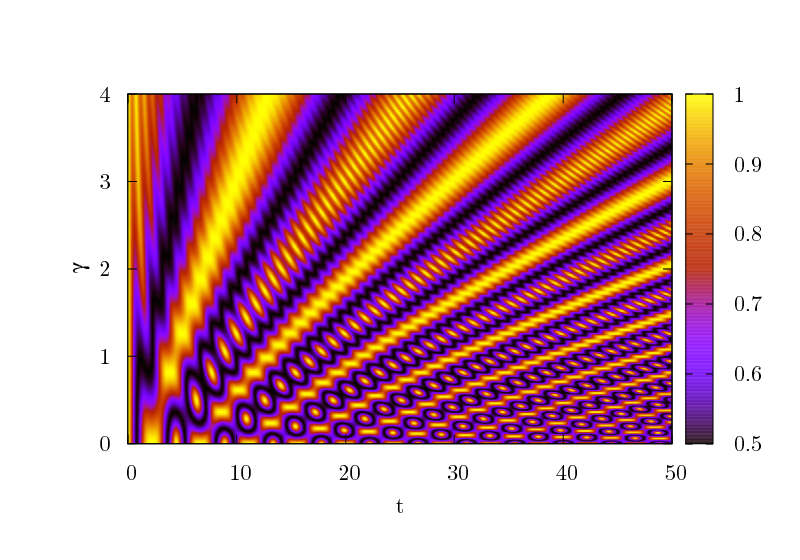}
  \caption{
  Purity of the field  as a function of time
  and the rest energy $\gamma=mc^2$,
   with an initial state formed as a product of the lowest DMO eigenstate $E=0$ and
  the upper state of the isospin field $\ket{+}$. $\chi=\eta$. The period
  of oscillations increase with $\gamma$ an effect already found in
  \cite{emerson} which holds only for the resonance $\gamma=mc^2$.
  }\label{fig:purity}
\end{figure}
Our next task is to find the entanglement with the field. To this end one has to evaluate
the reduced density matrix of the field by taking a partial trace over the DMO degrees of freedom.
One finds the density matrix
\begin{align}
  \rho'(t)=\Tr_{{\rm DMO}}
  \left\{\ket{\Psi(t)}\bra{\Psi(t)}\right\}=
  \left(
  \begin{array}{cc}
\left|B_1(t)\right|^2+
\left|B_3(t)\right|^2&0\\
0&\left|B_2(t)\right|^2
  \end{array}\right)
\end{align}
The entanglement of the DMO with the field can be measured by the purity \cite{nielsen},
obtained as
\begin{equation}
  P_{ {\rm F} }=\Tr\{\rho'^2\}=\tfrac{1}{2}+\tfrac{1}{2}
  \left(g(t)-f_0(t)
  \right)^2
\end{equation}

Figure \ref{fig:purity} shows the purity of the field as a function of $t$ and 
$\gamma=mc^2$, the field strength equal to the rest mass energy in this case.
The purity starts at value $P_{ {\rm F} }=1$ as expected for an initial product state. 
Full entanglement of the isospin with the D     occurs when purity
reaches it minimum value of $P_{ {\rm F} }=1/2$. Note  that the
oscillations between fully entangled and totally
pure situations have a period which increases with $\gamma$. 
Normally one could expect an increase of the frequency 
with $\gamma$, but as already observed in  
\cite{emerson}, this is an effect of the resonance $\gamma=mc^2$.

\section{Two atoms inside a cavity}\label{sec:cav2at}
In this section we explain the connection of the coupled DMO of section 
\ref{sec:DMOcoup} with quantum optical systems.
The Hamiltonian in equation \eqref{eq:hamDOMc}
can also be used in the context of quantum optics to describe a system
composed of two atoms inside
a cavity, provided one identifies $\mathcal{A}$ with $a$ the ladder
operator of the cavity and each isospin with an atom.
Besides one has to regard $\eta$ and $\chi$ as the coupling
of each atom to the cavity and $mc^2$ and $\gamma$ as the detuning
of each transition level with the cavity mode.

This means that the extended model can also be 
mapped in cavity QED. Again the $1+1$ and $2+1$ cases can be mapped
exactly, whereas  the $3+1$ case can only be reproduced for the special situation
considered in subsection \ref{subsec:ef}, because
in $3+1$ dimensions the coupling with the field mixes the dynamics
of the infinitely degenerate part with the finitely degenerate part.
This case $f(N)$ in the blocks \eqref{eq:matDMOc} of $\tilde H$ 
to depend on $\mu(N)$ which changes by steps of $2j+2$
as $N$ increases. For $N=0$ the equation \eqref{eq:matDMOc} reduces
to a $3\time 3$ matrix and only the value $\mu(0)=2n$ comes into play.
Thus the particular case studied in subsection \ref{subsec:ef},
which can represent the three dimensionalities,
can be mapped to a quantum optical system. For this reason, in this work
we focus  only on the equivalent quantum optics model restricted
to $n=0$, that is an initially empty cavity. 

In the context of quantum information theory, this is one of the simplest models 
which can be used in order to study two important aspects: entanglement and decoherence. 
The former is a resource for implementing quantum information protocols - and therefore, 
our ally - while the latter is an obstacle for such implementation. While these 
concepts play an important role in nurturing the {\it quantumness \ } of a central 
system, here we shall make use of quantum information measures as a tool to infer 
dynamical features of our problem in the simplest possible way.

We shall not rewrite the Hamiltonian as it is equal to the one in equation 
\eqref{eq:hamDOMc} and
the same solutions apply. The initial state to consider here has to be different
as the quantities of interest are others. We distinguish between the two atoms 
(central system) and
the cavity (environment). So we shall begin with a product state of 
an atomic state times a number state of the cavity, and for simplicity we again 
take $\mathcal{N}=0$
\begin{equation}\label{eq:inat}
  \ket{\Psi_0}=  \left(\cos{(\alpha)}\ket{-+}+\sin{(\alpha)}\ket{+-}\right)\ket{0}.
\end{equation}
note that for $\alpha=0$ the state is exactly the same as in \ref{eq:DMOcist}.
We shall consider equivalent conditions as in subsection \ref{subsec:ef}, which
in this case means equal couplings to the cavity and atomic transition
frequencies equally detuned from the cavity mode. With these considerations 
at time $t$ the state vector can be found in in a superposition like
in equation \eqref{eq:DMOctst}, the only difference is that the coefficients
will depend on $\alpha$, so we shall write them as
\begin{align}
C_1(t)=&f_\alpha(t)\nonumber\\
C_l(t)=&\tfrac{1}{2}\left(1-f_\alpha(t)+(-1)^l g(t)\cos{(2\alpha)}\right),
\quad l=2,3
\end{align}
where $g(t)$ and $f_0(t)$ are given in equation \eqref{eq:fg} and we introduced
\begin{equation}
  f_\alpha(t)=\left(1+\sin{(2\alpha)}\right)f_0(t).
\end{equation}
One has to remember that here $\gamma$ represents the detuning of both atoms.

\subsection{Entanglement measures}
To evaluate the measures of entanglement one needs of the reduced 
density matrix of the two atoms, so
we trace over the oscillator degrees of freedom to get:
\begin{equation}
  \rho=\Tr_{\rm Osc}\left\{\ket{\Psi(t)}\bra{\Psi(t)}\right\}
  =\left(
  \begin{array}{cccc}
    |C_1(t)|^2&0&0&0\\
    0&|C_2(t)|^2&\left(C(t)_3\right)^*C_2(t)&0\\
    0&\left(C(t)_2\right)^*C_3(t)&|C_3(t)|^2&0\\
    0&0&0&0
  \end{array}
  \right).
\end{equation}
To measure the entanglement between the two atoms (central system)
and the cavity (environment) we use the purity $P=\Tr\left\{\rho^2\right\}$
and find 
\begin{equation}
P(t)=1-2f_\alpha(t)+2f_\alpha^2(t)
  \label{eq:pur}
\end{equation}
which serves as measure  of the decoherence of the two atoms system. 

To measure the entanglement between the atoms we use the concurrence \cite{wootters}
$C(\rho)={\rm Max}\left\{0,\lambda_1-\lambda_2-\lambda_3-\lambda_4\right\}$,
where $\lambda_j$ are the eigenvalues of
$\left(\rho\,\sigma_y\sigma_y'\,\rho^*\,
\sigma_y\sigma_y'\right)^{1/2}$ in non-increasing order.
In this case we find
\begin{align}
  C(t)=\sqrt{\left(f_\alpha(t)-1\right)-g^2(t)\cos^2{(2\alpha)}}
\end{align}

A useful way of visualizing both dynamics together is the so called $CP$ plane 
\cite{buzek,torres}. 
Figure \ref{fig:cp} shows this plane for two separate cases,
both with detuning $\gamma=1$.
The left part shows in red for an initial state with $\alpha=0$,
the equivalent case to subsection \ref{subsec:ef}, whereas 
the right part shows it in red for $\alpha=\pi/40$.
On both sides the red curves are parametrized up to $t=30$.
The black curve on both figures shows the behavior
for zero detuning for $\alpha=\pi/4$ and also on the right
part for $\alpha=\pi/40$ (lower black curve).
Both figures show a gray area that correspond to states
with density matrices which are not physically acceptable and
its lower frontier corresponds to
the {\it maximally entangled mixed states}.

It is interesting to note
that in both cases the red curves form
Lissajous like behavior in the $CP$ plane. 
For zero detuning
analytic expressions for the concurrence as a function of the purity 
can be found 
\begin{eqnarray}
  C_{\pm}(P;\alpha)&=
  \frac{1}{2}\left|1\pm\sqrt{1+2(P-1)}-2\sin{(2\alpha)}\right|
\end{eqnarray}
these curves form the frontier of the region that the red curve 
partially fill as one can see in the figure \ref{fig:cp}. 
The region filled by the red curves reduces
as one increases the detuning $\gamma$, that is
the atoms feel less the effects of the cavity. 

We also found similar behavior  
in \cite{torres} when instead we had zero detuning
but interaction between the atoms.

\begin{figure}
  \includegraphics[width=.5\textwidth]{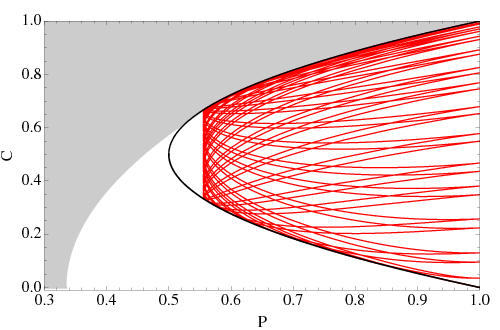}
  \includegraphics[width=.5\textwidth]{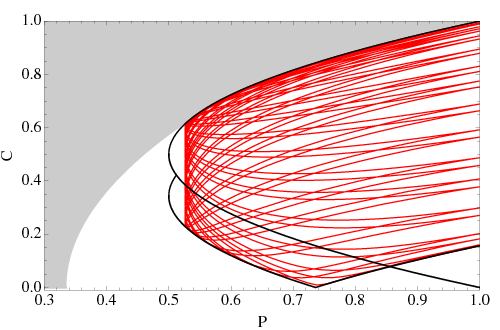}
  \caption{
  Concurrence versus purity plane for an initially empty cavity and
  two equal atoms.
  The black curves show the case with zero detuning $\gamma=0$
  and initial state $\alpha=\pi/4$ (see eq. \eqref{eq:inat}),
  ($\alpha=\pi/40$ bottom curve on the right.)
  In red the curves for detuning $\gamma=1$ for
  $\alpha=0$ on the left and $\alpha=\pi/40$ on the right.
  The gray area corresponds to unphysical density matrices and
  its lower frontier represents the so called
  {\it maximally entangled mixed states}. }\label{fig:cp0}
\end{figure}\label{fig:cp}
\section{Conclusions and final remarks}\label{sec:conclusions}

The Dirac Moshinsky oscillator 
in $1+1$ and $2+1$ dimensions was mapped to the Jaynes-Cummings model. 
For the $3+1$ case 
at least the infinitely degenerate part can be mapped. This scheme was 
based on explicitly using an invariant of these systems, which in    
the optical image 
acquires the simple meaning of the total 
number of excitations, which differs from the total 
energy because of detuning.

Based on this picture we obtained  a soluble extension to the DMO 
coupled to an isospin field, if this coupling is carefully
chosen to have an invariant that includes the excitations of the isospin field.
The system not only  retains solvability, but  can be mapped
to 2 atoms inside a cavity again for the $1+1$ and $2+1$ dimensional
cases. 

In the 2+1 dimensional DMO the presence of an external field which allows integrability
does not break the infinite degeneracy. The absence of the operator $n_l$ for the JC  
model can be thought as "inert" field modes which do not interact with the atoms in our 
system. In the quantum-optical system, other cavity modes may be populated by photons.
However, if their frequencies are far off-resonance with our two-level atoms, 
the Hamiltonian in the interaction picture will contain no terms related to such modes. 
To complete the analogy, a Dirac particle in 2+1 dimensions contains such degrees of 
freedom, but they have no effect on the energies of the DMO.

We recognized that in our setting for $3+1$ dimensions, only the block 
$N=0$ can be represented by this diatomic model. 
However, the additional structure 
in the relativistic model may lead to other soluble models in atom optics, 
known or unknown at present. This will have to be the subject of further investigations.

By way of example we evaluated the purity of the field by tracing
over the DMO degrees of freedom as well as other examples of
the entanglement measures of the quantum optical system to which we mapped 
many of the discussed models.


\begin{thebibliography}{99}
\bibitem{marcosbook} M. Moshinsky and Y. Smirnov,
  {\it The Harmonic Oscillator in Modern Physics\ }, 
  Hardwood Academic Publishers, Amsterdam, 1996.

\bibitem{mosh}M. Moshinsky and A. Szczepaniak,
  {\it J. Phys. A} {\bf 22}, L817 (1989).

\bibitem{lista}   
M. Moshinsky, {\it et. al.} {\it The two body Dirac oscillator}, AnniversaryVolume in Honor of J.J. Giambiagi, (World Scientific Press, Singapore, 1990).
M. Moshinsky, {\it et. al.} 
{\it Proceedings of the Rio de Janeiro International Workshop 
on Relativistic Aspects of Nuclear Physics}, 271-307 
(World Scientific, Singapore, 1990)
M. Moshinsky, {\it et. al.}
{\it Proceedings of the 13th Oaxtepec 
Symposium on Nuclear Physics}, Vol 13, No. 1 187-195 (1990).
M. Moshinsky, {\it et. al. } 
{\it Relativistic invariance of a many body system with a Dirac oscillator interaction}, 
Lecture Notes in Physics, 1991, Volume 382 (1991).

\bibitem{torres} 
  J.M. Torres, E. Sadurni and T.H. Seligman, 
  { \it J. Phys. A} {\bf 43}  192002 (2010)

\bibitem{lamata} 
  L. Lamata, J. Leon, T. Schaetz, E. Solano,
  {\it Phys. Rev. Lett.} {\bf 98}, 253005 (2007).

\bibitem{blatt}
  
  R. Gerritsma, G. Kirchmair, F. Z\"ahringer, E. Solano, R.
  Blatt, and C. F. Roos, {\it Nature} {\bf 463}, 68 (2010).

\bibitem{delgado} Bermudez A, Martin-Delgado M A and Solano E 2007,{\it Phys. Rev. A} {\bf 76} 041801.

\bibitem{bunch} Jaynes E T and Cummings F W 1963,{\it Proc. IEEE} {\bf 51} 89.

\bibitem{delgadochiral}  A. Bermudez, M. A. Martin-Delgado, and A. Luis {\it Phys. Rev. A} {\bf 77} 063815 (2008)



\bibitem{emerson}E. Sadurni, J.M. Torres and T.H. Seligman, 
  { \it J. Phys. A} {\bf 43} 285204 (2010)


\bibitem{nielsen} M. Nielsen and I. Chuang 
  {\it Quantum Computation and Quantum Information },
Cambridge University Press, Cambridge, 2000.

\bibitem{wootters} W. K. Wootters, 
  { \it Phys. Rev. Lett.} {\bf 80}, 2245 (1998).

\bibitem{buzek} M. Ziman and V. Bu\v{z}ek,
  {\it Phys. Rev. A.} {\bf 72}, 052325 (2005).



\end{thebibliography}
\end{document}